\documentclass[dvips]{acta}
\usepackage{supertabular,lscape,epsfig}
\usepackage{amssymb}
\usepackage{amsmath}
\SetPages{1}{11}
\SetVol{59}{2009}

\begin{document}

\begin{Titlepage}

\Title { New Interpretation of Superhumps }

\Author {J.~~S m a k}
{N. Copernicus Astronomical Center, Polish Academy of Sciences,\\
Bartycka 18, 00-716 Warsaw, Poland\\
e-mail: jis@camk.edu.pl }

\Received{  }

\end{Titlepage}

\Abstract { 
Observational evidence is presented for periodically variable irradiation of 
secondary components. This results in strongly modulated mass outflow. 
Superhumps are then due to enhanced dissipation of the kinetic energy of the stream. 

Qualitative interpretation of superhump periods and their variations is also 
presented. } 
{accretion, accretion disks -- binaries: cataclysmic variables, stars: dwarf novae }

\section {Introduction }

Superhumps are common among dwarf novae of the SU UMa subtype during 
their superoubursts (Warner 1995, Hellier 2001). They are also observed 
in the case of the so-called permanent superhumpers (cf. Patterson 1999). 
The superhump periods are slightly longer than the orbital periods and their 
amplitudes are -- typically -- 0.3 mag. 

The commonly accepted "tidal" model explains superhumps as being due to tidal 
effects in the outer parts of accretion disks leading -- via the 3:1 resonance 
-- to the formation of an eccentric outer ring undergoing apsidal motion. 
This model and, in particular, the results of numerous 2D and 3D SPH simulations 
(cf. Smith et al. 2007 and references therein) successfully reproduce the observed 
superhump periods and correlations of the superhump period excess with the orbital 
period and the mass-ratio.  
Moreover, those correlations find natural interpretation in the context of structure 
and evolution of the secondary components (Patterson 1998,2001, Patterson et al. 2005). 

On the other hand, however, the tidal model fails to reproduce the amplitudes 
of superhumps: The numerical SPH simulations produce "superhumps" which have 
amplitudes $\sim 10$ times too low, compared to the observed amplitudes (Smak 2009a). 
This implies that the nature of superhumps requires another explanation. 

Recent analysis of superhumps in Z Cha (Smak 2009b) showed that -- contrary to 
the "tidal" interpretation -- they are due to strongly modulated mass transfer rate. 
The question then appears: what is the origin of those modulations? 
The answer to this question is given in the present paper. 

In Section 2 we present observational evidence for variable irradiation of 
secondary components. Its consequences are discussed in Section 3 in terms of the irradiation modulated mass outflow.   
In Section 4 we discuss the relationship between the irradiation periods and the 
observed superhump periods and present general discussion of their variability. 
Results are summarized in Section 5.

\section { The Missing Link }

Many years ago Schoembs (1986) discovered that the luminosity of OY Car 
during its November 1980 superoutburst varied with period close to the 
beat period. Until now, however, the significance of this discovery has not been 
fully appreciated. 

The see whether such a behavior is common among other dwarf novae a search was made 
through the literature for light curves well covering all beat phases. 
Fifteen such cases were found, including 12 dwarf novae observed during their 
superoutbursts and 3 permanent {\it positive} superhumpers. 
They are listed in Table 1, where the first six stars, which are eclipsing systems, 
are arranged in order of decreasing orbital inclinations, while the remaining, non-eclipsing systems -- in alphabetical order. 
For eight of them it was possible to use the pure disk magnitudes measured away 
from eclipses and from superhumps. For the others -- only nightly mean 
magnitudes were available.

\begin{table}[h!]
{\parskip=0truept
\baselineskip=0pt {
\medskip
\centerline{Table 1}
\medskip
\centerline{ Photometric Data }
\medskip
$$\offinterlineskip \tabskip=0pt
\vbox {\halign {\strut
\vrule width 0.5truemm #&	
\enskip\hfil#\enskip&	        
\vrule#&			
\enskip\hfil#\hfil\enskip&      
\vrule#&			
\enskip#\hfil\enskip&	        
\vrule#&			
\enskip#\hfil\enskip&	        
\vrule width 0.5 truemm # \cr	
\noalign {\hrule height 0.5truemm}
&&&&&&&&\cr
&Star\hfil&& $i$ &&\hfil Data source && &\cr
&&&&&&&&\cr
\noalign {\hrule height 0.5truemm}
&&&&&&&&\cr
&  IY UMa  &&86.8&&Patterson et al. (2000a), Fig.2, JD2451562-569  &&2   &\cr
&  DV UMa  &&84.0&&Patterson et al. (2000b), Table 1 + Fig.3, 1997 &&3   &\cr
&  OY Car  &&83.3&&Schoembs (1986), Fig.1                          &&2   &\cr
&   Z Cha  &&80.2&&Kuulkers et al. (1991), Table 3                 &&3   &\cr
&  WZ Sge  &&75.9&&Patterson et al. (2002), Table 1, JD2452126-137 &&3   &\cr
&   U Gem  &&69.0&&Smak and Waagen (2004), JD 2446344-376          &&3,4 &\cr
&V603 Aql  &&    &&Patterson et al. (1993a), Table 1               &&3   &\cr
&  VY Aqr  &&    &&Patterson et al. (1993b), Fig.7, JD2446558-566  &&2,5 &\cr
&  TT Ari  &&    &&Wu et al. (2002), Fig.1                         &&2   &\cr
&  TT Boo  &&    &&Olech et al. (2004), Fig.3                      &&2,5 &\cr
&V503 Cyg  &&    &&Harvey et al. (1995), Table 3, JD2449601-609    &&3   &\cr
&  VW Hyi  &&    &&Schoembs and Vogt (1980), Fig.2a, JD2443808-815 &&2   &\cr
&  BK Lyn  &&    &&Skillman and Patterson (1993), Table 1, 1992/93 &&1,3 &\cr
&  TU Mon  &&    &&Stolz and Schoembs (1984), Fig.2                &&2   &\cr
&  KS UMa  &&    &&Olech et al. (2003), Fig.3                      &&2,5 &\cr
&&&&&&&&\cr
\noalign {\hrule height 0.5truemm}
}}$$

\parindent=0 truemm
\parskip=3truept

Remarks: 1. PG 0917+342,  2. pure disk magnitudes, 3. mean magnitudes, 
4. mean points in 0.1 phase bins, 5. zero-point of $\phi_b$ arbitrary.  

\parindent=12 truemm
\parskip=12truept
}}
\end{table}

We analyze those light curves in the following way. 
To begin with, using the orbital and superhump elements, we calculate the 
beat phases 

\beq
\phi_b~=~\phi_{orb}~-~\phi_{sh}~.
\eeq 

\noindent
In three cases, where no orbital periods were available, their values had to be 
determined from their superhump periods using formula given by Menninckent et al. 
(1999). Consequently an arbitrary zero-point of $\phi_b$ had to be adopted.  

Anticipating further considerations it may be worth recalling that the beat phase 
refers to the beat period which is defined by the other two periods as 

\beq
{1\over{P_b}}~=~{1\over{P_{orb}}}~-~{1\over{P_{sh}}}~. 
\eeq 

The observed light curves are fitted with a simple formula:  

\beq
m~=~m~(t_\circ)~+~{{dm}\over{dt}}~(t-t_\circ)~-~A~\cos~(\phi_b-\phi_b^{max})~, 
\eeq 

\noindent
where the first two terms describe the slow brightness decline during superoutburst, 
while the third -- the expected modulation with the beat period. 
The residual magnitudes are then determined as

\beq
\Delta m~=~m~-~\left [~m~(t_\circ)~+~{{dm}\over{dt}}~(t-t_\circ)~\right ]~.  
\eeq 

Results are listed in Table 2 and shown in Figs.1 and 2. 
As we can see there is an obvious dependence on the orbital inclination: 
(1) There are only three cases showing large variations of $\Delta m$ with $\phi_b$, 
with full amplitudes $2A\sim 0.35-0.40$ mag. Those are the deep eclipsers with 
highest orbital inclinations (OY Car, IY UMa and DV UMa). Worth noting is that they 
have also very similar phases of maximum. 
(2) The other three eclipsing systems with lower inclinations (Z Cha, WZ Sge and 
U Gem), show much smaller amplitudes which, in view of their large errors and large 
scatter of points seen in Fig.1, can hardly be considered significant. 
Plotted in Fig.2 is also another eclipsing dwarf nova XZ Eri, which was 
analyzed earlier by Uemura et al. (2004). During one of its superoutbursts they 
detected low amplitude variations with a period $P=4.7$d, but no measurable 
variations with the beat period $P_b=2.33$d. It is plotted in Fig.2 with $A=0$
at $i=80.2$ as an asterisk. 
(3) The non-eclipsing systems, with low inclinations, do not show any significant 
variations.

\begin{table}[h!]
{\parskip=0truept
\baselineskip=0pt {
\medskip
\centerline{Table 2}
\medskip
\centerline{ Results }
\medskip
$$\offinterlineskip \tabskip=0pt
\vbox {\halign {\strut
\vrule width 0.5truemm #&	
\enskip\hfil#\enskip&	        
\vrule#&			
\enskip\hfil#\hfil\enskip&	
\vrule#&			
\enskip\hfil#\hfil\enskip&	
\vrule width 0.5truemm #&	
\enskip\hfil#\enskip&	        
\vrule#&			
\enskip\hfil#\hfil\enskip&	
\vrule#&			
\enskip\hfil#\hfil\enskip&	
\vrule width 0.5 truemm # \cr	
\noalign {\hrule height 0.5truemm}
&&&&&&&&&&&&\cr
&Star\hfil&&$A({\rm mag})$&&$\phi_b^{max}$&&Star\hfil&&$A({\rm mag})$&&$\phi_b^{max}$&\cr
&&&&&&&&&&&&\cr
\noalign {\hrule height 0.5truemm}
&&&&&&&&&&&&\cr
&  IY UMa&&0.18$\pm$0.04&&0.67$\pm$0.04&&  VY Aqr&&0.02$\pm$0.05&&    .....    &\cr 
&  DV UMa&&0.17$\pm$0.06&&0.60$\pm$0.06&&  TT Ari&&0.02$\pm$0.02&&0.38$\pm$0.12&\cr
&  OY Car&&0.20$\pm$0.07&&0.65$\pm$0.07&&  TT Boo&&0.05$\pm$0.03&&    .....    &\cr
&   Z Cha&&0.06$\pm$0.09&&0.28$\pm$0.25&&V503 Cyg&&0.02$\pm$0.06&&0.10$\pm$0.21&\cr
&  WZ Sge&&0.03$\pm$0.03&&0.49$\pm$0.13&&  VW Hyi&&0.05$\pm$0.03&&0.20$\pm$0.20&\cr
&   U Gem&&0.04$\pm$0.04&&0.35$\pm$0.18&&  BK Lyn&&0.02$\pm$0.01&&0.84$\pm$0.17&\cr
&        &&  &&  		       &&  TU Mon&&0.01$\pm$0.08&&0.59$\pm$0.23&\cr
&V603 Aql&&0.02$\pm$0.02&&0.35$\pm$0.19&&  KS UMa&&0.02$\pm$0.01&&    .....    &\cr
&&&&&&&&&&&&\cr
\noalign {\hrule height 0.5truemm}
}}$$
}}
\end{table}

\begin{figure}[htb]
\epsfysize=18.5cm 
\hspace{0.1cm}
\epsfbox{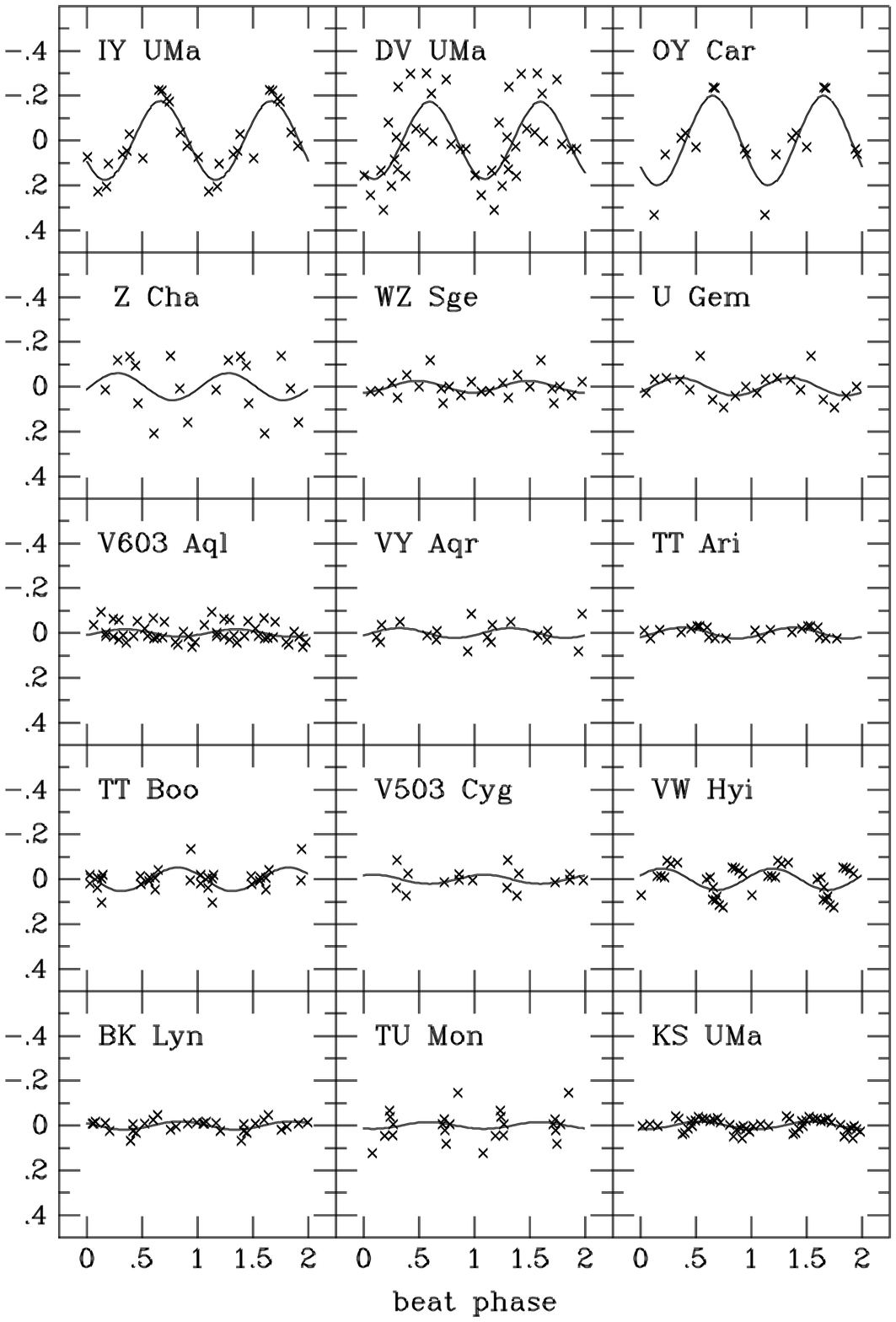} 
\vskip 5truemm
\FigCap { Residual magnitudes versus beat phase for stars listed 
in Table 2. Solid lines are cosine curves with $A$ and $\phi_b^{max}$ obtained 
from solutions with Eq.(3). }
\end{figure}

\begin{figure}[htb]
\epsfysize=8.0cm 
\hspace{1.5cm}
\epsfbox{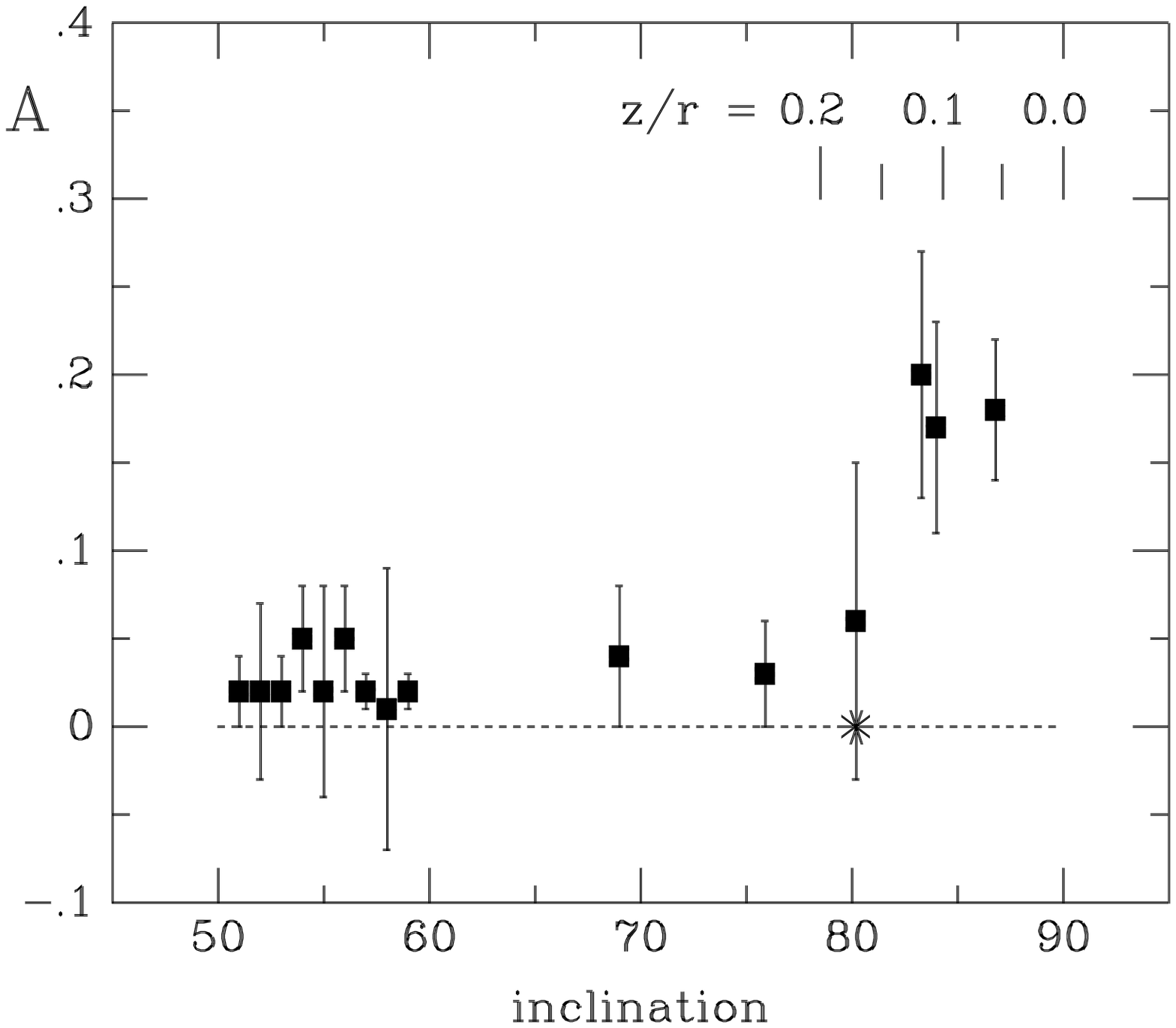} 
\vskip 5truemm
\FigCap {  The amplitude as a function of inclination for stars listed in Table 2. 
Non-eclipsing systems with unknown inclinations are plotted between $i=50^\circ$ and 
$60^\circ$. An asterisk denotes XZ Eri. Marked above are values of $z/r=\cos i$. } 
\end{figure}

There appears to be a simple interpretation of those results. 
Disks in dwarf novae during their superoutbursts and in permanent superhumpers 
have strongly non-axisymmetric structure, rotating -- in the observer's frame -- 
with a period $P_{rot}$, which is related, but only approximately equal 
(see Section 4), to the observed beat period $P_b$.  
In particular, the effective geometrical thickness of the disk $z/r$ must depend 
on the position angle or -- in the observer's non-rotating frame -- on the phase 
of $P_{rot}$.  
If so, the observed periodic light variations with period $P_b$ can be interpreted 
as being due to variable obscuration of the central parts of the disk, resulting 
from variable $z/r$. 

It is obvious that such variations can be seen only at orbital inclinations 
fulfilling the condition: $\cos i \approx z/r$. 
As can be seen from Fig.2, the three systems showing such variations have orbital 
inclinations corresponding to $z/r=0.06-0.12$. This is, indeed, typical for hot 
disks with $\dot M=10^{17}-10^{18}$g/s (cf. Smak 1992). 
In the case of Z Cha and XZ Eri, however, we get $z/r=0.17$ which would require unacceptably high accretion rate $\dot M=10^{19}$g/s. 

Modulations of disk luminosities with $P_{rot}$ (in the non-rotating 
frame of the observer) imply variable irradiation of the secondary component 
(in its rotating frame) with "irradiation period" $P_{irr}$ which -- in analogy 
to Eq.(2) -- is defined by 

\beq
{1\over{P_{rot}}}~=~{1\over{P_{orb}}}~-~{1\over{P_{irr}}}~,   
\eeq 

\noindent 
and therefore is approximately (see Section 4) equal to the observed superhump period.

\section { Irradiation Modulated Mass Outflow }

Irradiation of the secondary component was discussed by many authors 
as an important factor controlling the mass outflow from the inner 
Lagrangian point L$_1$. The main problem here is connected with the fact 
that the equatorial parts of the secondary, including L$_1$, are in the shadow 
cast by the disk and therefore are not directly affected by irradiation. 
Results of very crude model calculations (Smak 2004) showed that 
(1) the hot material flowing from irradiated regions can reach the vicinity 
of L$_1$, and (2) its temperature at L$_1$ is still high enough to produce  
substantial enhancement in the mass outflow rate. 
More recent model calculations by Viallet and Hameury (2007) confirmed 
the first conclusion, but not the second: according to them the flow, when 
it reaches L$_1$, is already too cool to produce any significant effects.  
It appears, however, that this result was mainly due to one of their assumptions. 
The rate of cooling of an isothermal layer is determined by its surface 
effective temperature $T_e$ which -- due to that cooling -- is lower than the 
temperature $T$ of the layer itself. Viallet and Hameury assumed $T_e=T$ thereby overestiming the cooling effects.   
Further model calculations are needed. 

The superhump phase of maximum irradiation of the secondary component 
$\phi_{sh}^{max}$ is related to the observed phase of maximum luminosity 
$\phi_b^{max}$ (see previous Section) by $\phi_{sh}^{max}=-~\phi_b^{max}$. 
This means that the maximum irradiation preceeds the next superhump by 

\beq
\Delta t~=~\phi_b^{max}~P_{sh}~=~\phi_b^{max}~
         { {P_{sh}}\over {P_{orb}} } ~P_{orb}~.
\eeq

\noindent
This delay consists of two terms: 

\beq
\Delta t~=~\Delta t_{flow}~+~\Delta t_{str}~, 
\eeq

\noindent
where $\Delta t_{flow}$ is the time needed for the flow to reach L$_1$, 
and $\Delta t_{str}$ is the time needed for the stream to reach 
the point of impact. 

The last equation can be used to estimate $\Delta t_{flow}$. 
From IY UMa, DV UMa and OY Car, using their average values of $<\phi_b^{max}>=0.64$ 
and $<P_{sh}/P_{orb}>=1.028$, we get $\Delta t=0.66~P_{orb}$. 
Values of $\Delta t_{str}$ can be obtained from stream trajectory calculations. 
With $<q>=0.13$ we get from $\Delta t_{str}=0.19~P_{orb}$ at the edge of the disk 
to $\Delta t_{str}=0.23~P_{orb}$ at a point half way between disk edge and its center. 
Inserting those values into Eq.(7) we obtain: $\Delta t_{flow}=(0.43-0.47)~P_{orb}$. 

Two comments are worth to make at this point. 
First, that surprisingly similar values: $\Delta t_{flow}\sim 0.4~P_{orb}$ 
were predicted by our crude model calculations (Smak 2004, Table 2). 
Secondly, that -- should the flow time be actually much longer -- then our result 
would have to be rewritten in a more general form as 
$\Delta t_{flow}=(0.43-0.47)~P_{orb}~+~n~P_{orb}$.

\section { The Superhump Periods and their Variations }

In the previous Sections we discussed only the effects of periodic variations 
of $z/r$ which occur on a short time scale corresponding to $P_{sh}$. 
Let us now consider its variations on a longer time scale. 
During superoutbursts the mass transfer and accretion rates change considerably 
resulting in significant variations of the irradiating flux and the effective 
geometrical thickness of the disk (averaged over all position angles). 
Generally we have: $F_{irr}\sim \dot M$ and $z/r\sim \dot M$ (cf. Smak 1992). 
Therefore: $d F_{irr}/d \dot M>0$ and $d (z/r)/d \dot M>0$. 
The flow time depends on both those factors and, in particular, we expect 
$\partial \Delta t_{flow}/\partial F_{irr}<0$ and 
$\partial \Delta t_{flow}/\partial (z/r)>0$. 
The flow time is then predicted to vary as  

\beq
{{d \Delta t_{flow}}\over {d t}}~=~\left [~
{{\partial \Delta t_{flow}}\over {\partial (z/r)}}~
{{d (z/r)}\over {d \dot M}}~+~
{{\partial \Delta t_{flow}}\over {\partial F_{irr}}}~
{{d F_{irr}}\over {d \dot M}}~
\right ]~{{d \dot M}\over {d t}}~. 
\eeq

\noindent 
We note that the first term in the square brackets (with $z/r$) 
is positive while the second term (with $F_{irr}$) is negative. 
Therefore, depending on which of the two factors is dominant, the expression 
in the square brackets can be either positive or negative. 

Taking into account variations of $\Delta t_{flow}$, and assuming that 
$\Delta t_{str}$ is practically constant, we obtain the following equation 
connecting the observed superhump period with irradiation period 
(introduced in Section 2 via Eq.5) 

\beq
P_{sh}~=~P_{irr}~\left [~1~+~{{d\Delta t_{flow}}\over {d t}}~\right ]~. 
\eeq

\noindent
As we can see, the two periods are equal only when $\Delta t_{flow}=$const 
(in such a case we also have $P_b=P_{rot}$). 
When $\Delta t_{flow}$ changes with time the two periods differ by a small amount, 
depending on the value and sign of $d\Delta t_{flow}/d t$. 

Let us now turn to the problem of superhump period variations. 
Up to mid 1990-ies nearly all dwarf novae observed until that time showed superhumps  
with periods which were {\it decreasing} during superoutburst (see Warner 1995). 
This was interpreted (cf. Lubow 1992) as being due to decreasing disk radius. 
Such an effect was indeed predicted by the thermal-tidal instability model 
of Osaki (1989). In particular, his calculations showed that the contraction 
of the disk begins at the very beginning of the superoutburst and continues throughout its main part. This interpretation, however, must be abandoned for 
two reasons. First, there is no observational evidence for such disk radius 
variations. Secondly, superoutbursts are not due to thermal-tidal instability, 
but due to a major enhancement in the mass transfer rate (Smak 2008). 
As a result, the accretion is quasi-stationary, with strongly enhanced accretion 
rate, causing disk radius to be close to the tidal radius and remain constant 
throughout the entire superoutburst. 

Situation became more complicated with the discovery of a group of dwarf 
novae with shortest orbital periods showing {\it increasing} superhump periods 
(Kato et al. 2001,2003, Olech et al. 2003, and references therein) and 
even more so with the discovery of several examples of more complex 
{\it alternating} period variations (e.g. Olech et al. 2004, Rutkowski et al. 2007). 

The new interpretation of superhumps emerging from considerations presented above 
adds another dimension to this problem. From Eq.(9) we get 

\beq
{{d P_{sh}}\over {d t}}~=~{{P_{sh}}\over {P_{irr}}}~{{d P_{irr}}\over {d t}}~
+~P_{irr}~{{d^2 \Delta t_{flow}}\over {d t^2}}~.   
\eeq

\noindent 
This equation shows that superhump periods can change due to 
(1) variations of $P_{irr}$ (resulting from variations of $P_{rot}$), and 
(2) non-linear variations of $\Delta t_{flow}$ (as described by its second 
derivative). It explains why the observed period variations can be so complex. 

At the present moment we can present -- mainly for illustrative purposes -- 
only some crude, qualitative considerations. 
For this purpose let us arbitrarily assume (1) that $P_{irr}$ is constant, and 
(2) that the expression in the square brackets in Eq.(8) remains constant or 
changes with time only slightly so that the main contribution to the second 
derivative of $\Delta t_{flow}$ in Eq.(10) comes from variations of $\dot M$.  
If so, Eq.(10) can be replaced with 

\beq
{{d P_{sh}}\over {d t}}~\approx~P_{irr}~
\left [~{{\partial \Delta t_{flow}}\over {\partial (z/r)}}~
{{d (z/r)}\over {d \dot M}}~+~
{{\partial \Delta t_{flow}}\over {\partial F_{irr}}}~
{{d F_{irr}}\over {d \dot M}}~\right ]~
{{d^2 \dot M}\over {d t^2}}~.   
\eeq

\noindent
The accretion rate decreases during the main part of the superoutburst.  
In particular, we have $d\log\dot M/dt\approx$const (cf. Fig.3 in Smak 2008) 
which implies $d^2\dot M/d t^2>0$. 
The resulting variations of $P_{sh}$, however, will be either positive or 
negative, depending on the sign of the expression in the square brackets:   
When $z/r$ is the dominant factor we predict $dP_{sh}/dt>0$.  
Conversely, when $F_{irr}$ is the dominant factor we expect $dP_{sh}/dt<0$. 

Turning to observations we recall that superhump periods of dwarf novae with 
longer periods are generally {\it decreasing}, while those of dwarf novae 
with shortest periods are {\it increasing} (Kato et al. 2003, Fig.14, 
Olech et al. 2003, Fig.9). The transition occurs around $P_{tr}\sim 0.06$d. 
Our predictions could then suggest that the dominant factor in the first case 
is $z/r$, while in the second -- $F_{irr}$. 
In addition, there is a group of dwarf novae showing {\it alternating} variations 
of $P_{sh}$ (Olech et al. 2003, 2004, Rutkowski et al. 2007). 
Their periods are close to $P_{tr}$ and this could suggest that in their case 
$z/r$ nd $F_{irr}$ are equally important, resulting in a more complex behavior. 

Needless to say, detailed model calculations will be needed to confirm this 
simple interpretation and to perform the quantitative analysis of the problem.

\section { Discussion: the Nature of Superhumps }

Combining earlier evidence showing that superhumps in Z Cha are due to strongly 
modulated mass transfer rate (Smak 2009b) with evidence presented in the present 
paper we propose the following new intepretation of superhumps:  

{\parskip=0truept {
(1) Irradiation of secondary components is observed to vary periodically  
with period $P_{irr}$ which is close (but not exactly equal) to the observed 
superhump period $P_{sh}$. The basic "clock" mechanism must then be the same 
as in the case of the "tidal" model. 

(2) Variable irradiation of the secondary component results in mass ouflow being 
strongly modulated with period equal to the observed superhump period. 

(3) Superhumps are then due to enhanced dissipation of the kinetic energy of 
the stream. 
}}
\parskip=12truept 

Two kinds of models are now needed for a more meaningful, quantitative interpretation 
of superhumps and, in particular, of their variable periods: 

{\parskip=0truept {
(1) Realistic 3D disk models, including the effects of stream overflow and variable 
mass transfer rate. They are needed ({\it i}) to identify the nature of the "clock" 
and ({\it ii}) to determine $z/r$ and $F_{irr}$ as functions of time.  

(2) Models describing the irradiation modulated mass outflow. 
They are needed ({\it i}) to determine $\Delta t_{flow}$ and $\dot M$ as functions 
of $z/r$ and $F_{irr}$ (see Section 4) and ({\it ii}) to study the relationship 
between $P_{irr}$ and $P_{sh}$. 
}}
\parskip=12truept 

It can be hoped that once such models become available it will be possible 
to discuss not only the {\it common} superhumps and their periods, but also 
the other members of the "superhump zoo" (cf. Patterson et al. 2002). 

\Acknow{The author is grateful to Dr. Arkadiusz Olech for helpful discussions. }

\begin {references}

\refitem {Harvey, D., Skillman, D.R., Patterson, J., Ringwald, F.A.}
         {1995} {\PASP} {107} {551} 

\refitem {Hellier, C.} {2001} {{\it Cataclysmic Variable Stars} (Springer)} {~} {~}

\refitem {Kato, T., Sekine, Y., Hirata, R.} {2001} 
         {\it Publ.Astr.Soc.Japan} {53} {1191}

\refitem {Kato, T. et al.} {2003} {\MNRAS} {339} {861}

\refitem {Kuulkers, E., van Amarongen, S., van Paradijs, J., R{\"o}ttering, H.} 
         {1991} {\AA} {252} {605}

\refitem {Lubow, S.H.} {1992} {\ApJ} {401} {317}

\refitem {Menninkent, R.E., Matsumoto, K., Arenas, J.} {1999} {\AA} {348} {466}

\refitem {Olech, A., Schwarzenberg-Czerny, A., K{\c e}dzierski, P. Z{\l}oczewski, K., 
         Mularczyk, K., Wi{\'s}niewski, M.} {2003} {\Acta} {53} {175} 

\refitem {Olech, A., Cook, L.M., Z{\l}oczewski, K., Mularczyl, K., 
         K{\c e}dzierski, P., Udalski, A., Wi{\'s}niewski, M.} 
         {2004} {\Acta} {54} {233} 

\refitem {Osaki, Y.} {1989} {\it Publ.Astr.Soc.Japan} {41} {1005}

\refitem {Patterson, J.} {1998} {\PASP} {110} {1132} 

\refitem {Patterson, J.} {1999} {{\it Disk Instabilities in Close Binary Systems}, 
         Eds. S.Mineshige and J.C.Wheeler (Tokyo: Universal Academy Press)} {~} {61}

\refitem {Patterson, J.} {2001} {\PASP} {113} {736}

\refitem {Patterson, J., Thomas, G., Skillman, D.R., Diaz, M.} 
         {1993a} {\ApJS} {86} {235} 

\refitem {Patterson, J., Bond, H.E., Grauer, A.D., Shafter, A.W., Mattei, J.A.} 
         {1993b} {\PASP} {105} {69} 

\refitem {Patterson, J.,Kemp, J., Jensen, L., Vanmunster, T., Skillman, D.R.,
	 Martin, B., Fried, R., Thorstensen, J.R.} {2000a} {\PASP} {112} {1567}

\refitem {Patterson, J., Vanmunster, T., Skillman, D.R., Jensen, L., 
         Stull, J., Martin, B., Cook, L.M., Kemp, J., Knigge, C.} 
         {2000b} {\PASP} {112} {1584} 

\refitem {Patterson, J. et al.} {2002} {\PASP} {114} {721}

\refitem {Patterson, J. et al.} {2005} {\PASP} {117} {1204}

\refitem {Rutkowski, A., Olech, A., Mularczyk, K., Boyd, D., Koff, R.,
         Wi{\'s}niewski, M.} {2007} {\Acta} {57} {267}

\refitem {Schoembs, R.} {1986} {\AA} {158} {233} 

\refitem {Schoembs, R., Vogt, N.} {1980} {\AA} {91} {25}

\refitem {Skillman, D.R., Patterson, J.} {1993} {\ApJ} {417} {298}

\refitem {Smak,J.} {1992} {\Acta} {42} {323}

\refitem {Smak,J.} {2004} {\Acta} {54} {181}

\refitem {Smak,J.} {2008} {\Acta} {58} {55}

\refitem {Smak,J.} {2009a} {\Acta} {59} {***}

\refitem {Smak,J.} {2009b} {\Acta} {59} {this issue}

\refitem {Smak, J., Waagen, E.O.} {2004} {\Acta} {54} {433} 

\refitem {Smith, A.J., Haswell, C.A., Murray, J.R., Truss, M.R., 
         Foulkes, S.B.} {2007} {\MNRAS} {378} {785}

\refitem {Stolz, B., Schoembs, R.} {1984} {\AA} {132} {187} 

\refitem {Uemura, M. et al.} {2004} {\it Publ.Astr.Soc.Japan} {56} {141}

\refitem {Viallet, M., Hameury, J.-M.} {2007} {\AA} {475} {597} 

\refitem {Warner, B.} {1995} { {\it Cataclysmic Variable Stars} 
         (Cambridge University Press)} {~} {~}

\refitem {Wu, X., Li, Z., Ding, Y., Zhang, Z., Li, Z.} {2002} {\ApJ} {569} {418}

\end {references}
\end{document}